\documentclass[fleqn,10pt]{wlscirep}
\usepackage[utf8]{inputenc}
\usepackage[T1]{fontenc}
\usepackage{amsmath}
\usepackage{physics}
\usepackage{bm}

\title{Emergent learning: neuromorphic photonic computing with accelerated training}

\author[1,*]{Sara Pe\~{n}a-Guti\'errez}
\author[3]{Giorgio Gosti}
\author[4]{Hongsheng Chen}
\author[1,2]{Giancarlo Ruocco}
\author[1,5]{Marco Leonetti}

\affil[1]{Center for Life Nano- \& Neuro-Science, Italian Institute of Technology, 00185 Rome, Italy}
\affil[2]{Department of Physics, University Sapienza, 00185 Rome, Italy}
\affil[3]{CINECA, 00185 Rome, Italy}
\affil[4]{State Key Laboratory of Extreme Photonics and Instrumentation, College of Information Science and Electronic Engineering, Zhejiang University, Hangzhou, China}
\affil[5]{Institute of Nanotechnology of the National Research Council of Italy, CNR-NANOTEC, Rome Unit, Piazzale A. Moro 5, I-00185, Rome, Italy }

\affil[*]{sara.penagutierrez@iit.it}

\begin{abstract}
Emergent learning transforms a disordered optical medium into a photonic device capable of storage, recognition, and classification of arbitrary memory patterns. First, we show that the intensity at the output of a multiply scattering system can be described by a dyadic matrix, the \textit{optical-synaptic matrix}, exhibiting the same form as a Hebbian synaptic matrix containing a single memory. Then, we employ emergent learning — an approach inspired by neuroscience — to exploit the vast dictionary of \textit{raw memories} inherently available within a disordered optical structure, thereby engineering the \textit{optical-synaptic matrix} to store a user-defined attractor, or \textit{tailored memory}. Importantly these photonic structures also works as an optical comparators providing an intensity-based measure of the degree of similitude between a query pattern and the stored pattern, realizing an hardware co-localization between memory and optical operator. Our system has an almost infinite hardware capacity of \textit{tailored memories}/ operators ($\mathcal{M} \sim 10^{60557}$), thus these  \textit{tailored memories} can be then employed as examples to build a classifier hardware based on intensity comparison without the need of additional digital transformation layers.
Remarkably, this  Photonic Emergent Learning platform is not only flexible and fabrication-free, but also relies primarily on analog processes, thus shifting  the computational burden of training from the digital layers to the optical domain reducing the  computational cost and enhancing performance. 
\end{abstract}
\begin{document}

\flushbottom
\maketitle
%
%
\thispagestyle{empty}

\section*{Extended abstract}
Big data processing presents major challenges due to the immense volume and complexity of modern datasets, which often exceed the capabilities of traditional digital computing architectures. To address this, methods like reservoir computing (RC) have been developed to improve data handling. RC is a computational framework where the main processing occurs within a fixed, fully connected system called the reservoir, which transforms input data into a high-dimensional space. The training process is simplified, focusing solely on a basic readout layer, making RC both efficient and straightforward to implement, especially on physical computing platforms.

Building on this, analog optical systems emerge as a promising real solution by leveraging the physical properties of light to perform highly parallel and energy-efficient computations at the speed of light. An example of this are recent developments in analog optical computers that demonstrate drastically faster and less power-consuming processing of complex optimization and classification tasks, outperforming their conventional digital counterparts.\cite{Microsoft2025}.

Systems exhibiting disorder at the microscopic-nanoscopic scale provide a wealth of naturally occurring disordered patterns, which can be exploited for computing in the reservoir architecture. Indeed, RC has been obtained in disparate physical systems, including nanowires, memristors, spintronics \cite{daniels2022reservoir,milano2022materia, hon2022numerical}, and with an electromagnetic field in the temporal domain \cite{larger2017high}.

In the field of optics, diffusive media have the capability to project an optical input onto a transmitted output composed of hundreds of thousands of optical modes, thus effectively performing a projection to a space with dimensionality extremely larger than the original input. This form of optical computing is remarkably advantageous: it is an analog computing approach that exploits self-assembled materials, thus requiring no fabrication effort.

Intrinsic high dimensionality and complete connectivity, together with lightning-fast optical propagation, allows to perform complex transformations within a few picoseconds without the need for sculpting or drawing waveguides or optical fibers, as light propagates through diffusive media. By mapping these transformations into useful optical operations, it is possible to realize RC that enables parallel and analog computation at very high speed and with energy efficiency \cite{Wang2025}.

The typical spatial photonic reservoir computing (SPRC) consists of \textit{i)} an adaptive optical layer, which spatially modulates a propagating optical wavefront and serves as the input to be processed; \textit{ii)} a scattering medium acting as the reservoir, performing random optical transformations and projecting the input into a higher-dimensional transmitted mode space; and \textit{iii)} a detector layer acting as the output layer, where the (eventual) weights are stored and the system’s response to the input is retrieved.

The seminal work from the group of Gigan and Co-workers\cite{Saade2016}, takes advantage of this architecture employing the strategy of ridge regression with speckles ($RRS$) . To realize the classification,  the examples are projected through a Digital Micromirror Device (DMD), or a spatial light modularotr in the input layer. After recording the response for each example, a set of weights is designed specifically to perform classification for any query. However, this method requires significant computational effort, particularly inversion of a large $O\times O$ matrix, where $O$ is the number of light modes in the detection layer and $P$ is the number of samples. The approach can potentially be further improved in efficiency by incorporating nonlinearity \cite{skalli2022photonic, Wang2023}.

Building on the same architectural foundation, diffractive neural networks \cite{Lin2018, qian2020performing} arise, where the reservoir is actively engineered, i.e. by determining the transmission matrix through preliminary optical field forward simulations, followed by fabricating a diffraction mask representing the corresponding scattering matrix. Despite its power, this method is computationally and fabrication-wise expensive.

Different approaches within the same framework use recurrent approaches \cite{Xia2024,Yildirim2024} that harness multiple light interactions in scattering cavities or multilayer diffractive modulators to enable nonlinear processing and high-dimensional feature extraction. While balancing high-order nonlinear processing with energy efficiency, these approaches still rely on digital layers for final decoding, limiting fully all-optical autonomy.

Current bottlenecks in computation are energy consumption, hardware cost, and temporal requirements for training. In this context, the photonic neuromorphic RC architecture offers a reliable alternative and is attracting intense research interest. One possibility is to implement an optical equivalent of error backpropagation in spatial photonic neural networks \cite{spall2025training,pai2022inference}; however, this approach comes at the price of an increased experimental complexity and requires direct control of the neural weights (scattering coefficients).

In this paper, we present a new strategy to realize a spatial photonic reservoir computer (SPRC) that can store (\textbf{write}) and retrieve (\textbf{read}) stored memory patterns and \textbf{recognize} and \textbf{classify} queries from a user. 
In particular, we

\begin{enumerate}[label=(\roman*)]
\item propose a refined theoretical model describing how scattering from a multiple scattering medium produces an intensity defined by a Hebbian matrix, referred to as an \textit{optical-synaptic matrix};

\item identify how every transmitted light mode contains a unique, binary and random memory pattern, called a \textit{raw memory};

\item exploit a set of of tools developed for neuroscience to engineer a user designed attractor, effectively writing a user-defined emergent \textit{tailored memory} by selection specific \textit{raw memories} based on intensity;


\item demonstrate how \textit{tailored memories} can be further organized into deeper structures (\textit{memory classes}) capable of classification by simple comparison of intensities;

\item characterize reading and writing,  query recognition and classification processes through both simulations and experiments;

\item demonstrate photonic advantage in terms of classification efficiency and performance (FLOPs required for training).
\end{enumerate}

Unlike previous approaches, Photonic Emergent Learning ($PhEL$) fully exploits the immense number of disordered output modes naturally present in multiple-scattering transmission experiments. It demonstrates how these random structures can be organized into programmable, useful optical operators capable of analog computation, offering clear advantages.


\section*{Introduction}
\subsection*{Photonic Emergent Learning: Foundations and Simulations} 
As typical SPRC \cite{Vandersande2017,Saade2016,Dong2020}, employing the same optical architecture of a wavefront shaping system \cite{Popoff2010, Dremeau2015, leonetti2023reference}. Light is coupled to a coherent input beam transmitted through an adaptive optical system, effectively encoding the information to be processed onto a propagating electromagnetic wavefront $\bm{E}_n$. The index $n$ of the input mode runs over the  $N$ available elements of the adaptive optical input layer.

After propagating through a disordered, strongly scattering, non-absorbing medium, the input is transformed into the transmitted wavefront $\bm{E}_o$, as described by the transmission matrix model of disordered media \cite{devaud2021speckle}:

\begin{equation}
     E^o =\sum_n {t}^o_n E_{n}
    \label{eq:General SPRC field}
\end{equation}

where ${t}^o_n$ are the elements of the transmission matrix that connect each input mode $n$ to every output mode $o$ (with $O$ representing the size of our reservoir output layer). 

By introducing a field-transmitted vector $\bm{f}^o$, (where ${f}^o_n=t^o_n {E}_n$ represents the complex field contributing to the output spatial mode $o$, generated by the \textit{$n$}-th input element of the spatial light modulator in the absence of modulation) and $S_n$ (denoting the multiplicative coefficient applied to that field by the spatial light modulator), the previous expression can be rewritten as 

\begin{equation}
E^o =\sum_n {f}^o_{n} S_n    \label{eq:Modulator SPRC field}
\end{equation}



where $f_n^o$ has the dimensions of a complex field, while $S_n$ is an adimensional complex number. 
The intensity at a given output $o$ can therefore be expressed as

\begin{equation}
    I^o(\bm{S}) =|\sum_n {f_n^o}S_n|^2=|\bm{f}^o \cdot  \bm{S}^\dagger |^2=\bm{S} \cdot (\bm{f}^o\otimes\bm{f}^{o\dagger})   \cdot \bm{S}^\dagger =\bm{S} \cdot \mathbb{F}^o  \cdot \bm{S}^\dagger
    \label{eq:SPRC Intensity nu}
\end{equation}

where we introduced a vectorial/matrix formalism: the field-transmitted vector $\bm{f}^o$ corresponding to the output mode $o$, the input vector $\bm{S}$ (representing the spatial light modulator state), and the \textit{optical-synaptic matrix} $\mathbb{F}^o=\bm{f}^o\otimes\bm{f}^{o\dagger}$ (the elements of the \textit{optical-synaptic matrix} being $\mathbb{F}^o_{n,k}=f^o_nf^{o\dagger}_k$), which results from the tensor (dyadic) product of the field-transmitted vector with its conjugate transpose, see \hyperref[sec:Methods]{Methods}. Equation \ref{eq:SPRC Intensity nu} provides several important insights:

\begin{enumerate}[label=(\roman*)]

    \item $\mathbb{F}^o$ is a dyadic matrix, structurally analogous to the synaptic matrix used in the Hebbian learning framework 
    \cite{novellino2009modeling}.
    Within the language of Hopfield neural networks \cite{hopfield1982neural}, $\mathbb{F}^o$ has the same shape of the connectivity (synaptic) matrix storing memory $\bm{f}^{o}$ and thus we call it \textit{optical-synaptic matrix} (see methods). 
     
    \item The intensity $I^o(\bm{S})=|\bm{f}^o \cdot  \bm{S}^\dagger |^2$ reaches its maximum when the $N$-dimensional vector $\bm{S}$ is aligned to the transmission vector $\bm{f}^{o}$, specifically $\bm{S}=\bm{f}^{o}$. 
    
    \item Being the intensity $I^o(\bm{S})=|\bm{f}^o \cdot  \bm{S}^\dagger |^2$  a scalar product, it is also a proxy of the degree of similitude between the query and the pattern $\bm{f}^o$ "stored" in mode $o$. 
\end{enumerate}

The result $(ii)$ is connected to "wavefront shaping". It is well known that there exists a unique modulation configuration  of field and amplitude on an spatial light modulator that maximizes intensity at a given output \cite{vellekoop2008universal}. In other words, the wavefront shaping optimization problem is convex and the input configuration maximizing $I^o$ is unique and corresponds to $\bm{f}^{o \dagger}$. Here, we introduce the idea to use light modes as memory elements "containing" the pattern $\bm{f}^{o}$. In fact, the pattern can be retrieved at any moment just by measuring the \textit{transmitted field} vector \cite{Popoff2010,Conkey2012,Dremeau2015}$\bm{f}^{o}$ (that is, \textbf{reading} the memory).
We refer to these memories stored into a light structure originated by the disordered medium as \textit{raw memories}: those are uncontrolled and disordered, and therefore do not contain useful information. We will show below how emergent learning, enables to construct \textit{tailored memories} out of the \textit{raw memories}, i.e. that is, how to exploit this originally unusable repository of fully random structures in a constructive fashion.

At the basis of our approach lies the fact that (see insight \textit{iii)} ) each light mode works as an optical comparator. Indeed, when $\bm{S}=\bm{f}^{o}$ is presented to the input layer (as a query), the mode $o$ responds with maximal intensity  (\textbf{recognizes} the input). This means that our approach provides at the same time a unique recognition and comparison  device co-localized with the memory storage. This is a key and distinctive feature that enhances the performance of the $PhEL$, resembling biological neural networks in which neurons perform both operations.

As explained in \hyperref[sec:Methods]{Methods}, when real binary (boolean) modulation ($\bm{S}_n \in \pm 1$) is employed on the input channel each light mode "contains a \textit{raw memory}" slightly different from  $\bm{f}^o$, which is a boolean vector $\overline{\bm{d}}^o = sign(\bm{d}^o)$ corresponding to the sign of the eigenvector $\bm{d}^o$ of the real part of the \textit{optical-synaptic matrix} $\mathbb{R}^o=real(\mathbb{F}^o)\sim \bm{d}^o \otimes \bm{d}^{o\dagger}$.

The emergent learning works taking advantage of this immense wealth of random patterns by merging  a subset $\Sigma$ (containing $M<O$ modes) of all available modes by summing their intensities into an \textit{aggregated mode}. The \textit{aggregated intensity} $I^\Sigma(\bm{S})$ of this higher level optical structure is driven by a tailored \textit{optical-synaptic matrix} $\mathbb{R}^\Sigma$ ``containing'' the \textit{tailored memory} $\bm{d}^\Sigma$. The core of $PhEL$ lies in the formula describing $I^\Sigma(\bm{S})$ which results from the (incoherent) sum of individual light modes  $\in \Sigma$:

\begin{equation}
    I^\Sigma(\bm{S}) =\sum_m^{1..M}I^{m}(\bm{S}) =\sum_m^{1..M} \bm{S} \cdot \mathbb{R}^m  \cdot \bm{S}^\dagger= \bm{S} \cdot \left(\sum_m^{1..M}\mathbb{R}^m  \right)\cdot \bm{S}^\dagger=\bm{S} \cdot \mathbb{R}^\Sigma  \cdot \bm{S}^\dagger
    \label{eq:SPRC Intensity Sigma}
\end{equation}

In other words, we are using a repository of light modes as a dictionary, from  wich to extract $M$ letters to build a set $\Sigma$ realizing a \textit{tailored optical-synaptic matrix}  $\mathbb{R}^\Sigma= \sum_m^{1..M}\mathbb{R}^m $ containing the requested pattern $\bm{d}^\Sigma \sim \textbf{S}^*$.

\begin{figure}[ht]
\centering
unique\includegraphics[width=0.7\linewidth]{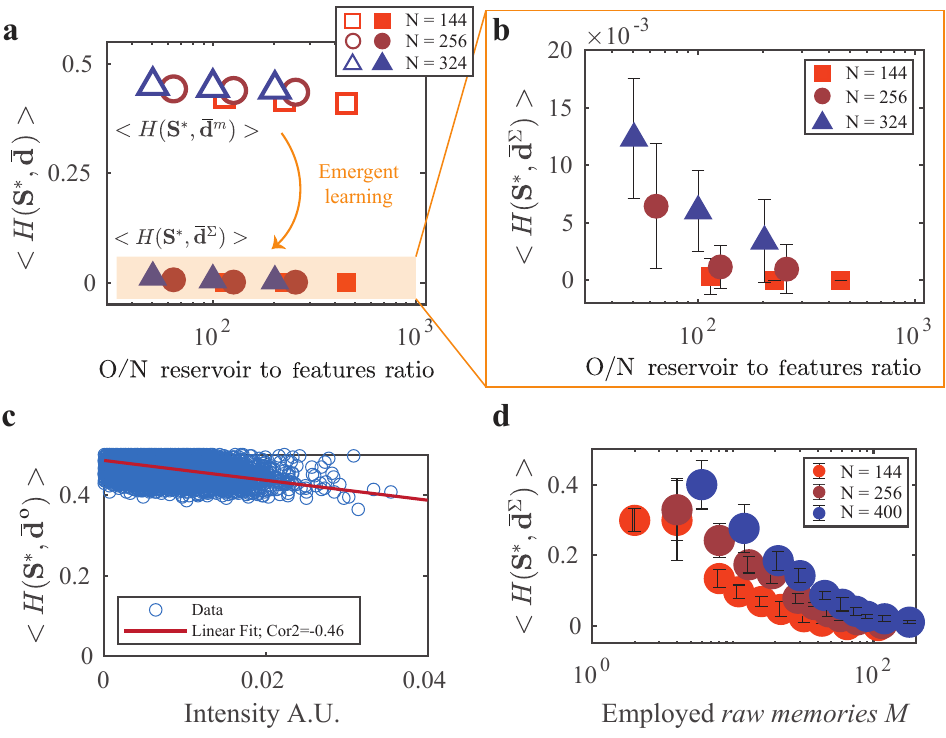}
\caption{(a) \textit{Demonstration of effectiveness of Emergence Learning}. Hamming distance of $\bm{S}^*$ from $\overline{\bm{d}}$ plotted against the ratio of $O/N$ reservoir to features ratio. Simulations performed with $N$ = 144, 256 and 324, $O$ = 500000 and using $M$ = 250 light modes. The memories used are the $m$th-selected modes $\overline{\bm{d}}^m$ more similar to the input and the final \textit{tailored memories} $\overline{\bm{d}}^\Sigma$ when aggregating these modes. (b) Detailed plot of the Hamming distance (a) in the case of using the \textit{tailored memories} $\overline{\bm{d}}^\Sigma$. (c) \textit{Intensity as a proxy of pattern distance} Hamming distance of $\overline{\bm{d}}^o$ from $\bm{S}^*$ plotted against intensity at the same mode. Simulations performed with $N$ = 144 and $O$ = 500000. (d)  \textit{Reading efficiency}. Fidelity of the retrieved-memory to the presented pattern measured through the Hamming distance for $O$ = 500000 using input patterns $\bm{S}^*$ of size N = 144, 256 and 400.}
\label{fig: Emergent Hamming}
\end{figure}

The operation of modes selection (the identification of the specific indices $o$ inside the big repository of $O$ modes) is the instrument we employ to effectively build the optical operator $\mathbb{R}^\Sigma$ in arbitrary fashion. To realize $\mathbb{R}^\Sigma \sim\bm{d}^\Sigma \otimes\bm{d}^{\Sigma\dagger}$ means to effectively \textit{write} the pattern $\bm{d}^\Sigma$ in our optical system and we can \textit{retrieve} the memory at any moment by reading the transmission vector $\bm{d}^\Sigma$. More importantly, as we stated before in point \textit{(iii)}, the aggregated mode works as an optical comparator, being capable to provide the bare measurement of the \textit{aggregated intensity}, the degree of similarity between an arbitrary query $\bm{S}$ and the stored pattern $\bm{d}^\Sigma$. We verified the validity of point \textit{(iii)} numerically. To realize the simulations, we first initialized $O=65536$ input light modes (complex vectors of $\bm{f}^o$ each with $N=18$ elements) and then calculated the intensity of each modes for a randomly generated input $\bm{S}^*$. Then, we calculated the the Hamming distance $H(\bm{S},\overline{\bm{d}}^o)$ (see \hyperref[sec:Methods]{Methods}) between the proposed patterns and the \textit{raw memory} stored in each mode. In Fig.\ref{fig: Emergent Hamming}c we report $H(\bm{S},\overline{\bm{d}}^o)$ versus $I(\bm{S}^o)$, showing a noisy but solid anti-correlation i.e. patterns providing higher intensity are closer to query (lower Hamming distance).  

Realizing this mode-selection operation is equivalent to \textit{write} and \textit{store} an arbitrary, user designed pattern $\bm{S}^*$ into our scattering-enabled memory. This process is a photonic generalization of the emergent archetype paradigm proposed by Agliari and coworkers \cite{Agliari2022} in the framework of numerically simulated Hopfield  networks, with the aim to install a specific memory/fixed point/attractor into a Hopfield network by summing noisy replicas of a noiseless target pattern thanks to a \textit{noise canceling} effect. In the photonic case, instead, we  leverage the immense repository provided by natural disorder in a scattering system, selecting the ones which have a good similarity (small Hamming Distance, see \hyperref[sec:Methods]{Methods}) from the desired pattern. In a nutshell, we seek to generate an \textit{aggregated mode} consisting of the sum of a set $\Sigma$ chosen such that $\mathbb{R}^\Sigma \sim\bm{d}^\Sigma \otimes\bm{d}^{\Sigma\dagger}$ with $\bm{d}^{o\in\Sigma}\sim \bm{S}^*$. 

\begin{figure}[ht]
\centering
\includegraphics[width=0.7\linewidth]{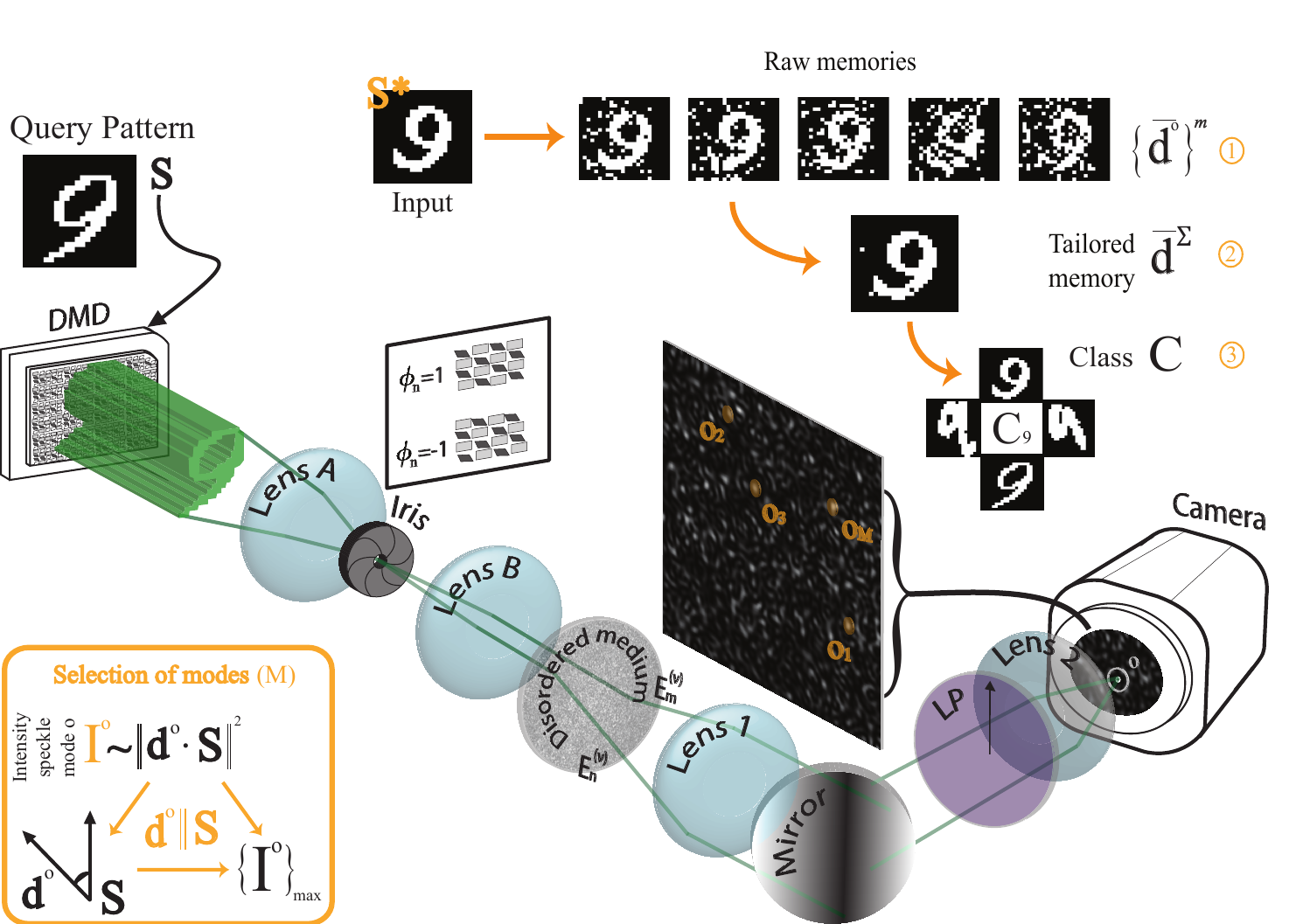}
\caption{\textit{Emergent reservoir optical scheme}. Illustration of the set-up and working conditions of the SPRC based on $PhEL$. The SPRC comprises a digital micromirror device, which projects the query pattern $\bm{S}$ into the disordered medium. Multiple scattered light is filtered by a linear polarizer (LP) and then it is focused onto the camera sensor. Inlet bottom left: Each mode of the speckle pattern $o$ has a correspondent \textit{raw memory} $\bm{d^{o}}$. The intensity of a speckle mode $I^o$ is related to the projection of \textit{raw memory} $\bm{d^{o}}$ onto the query pattern $\bm{S}$. The angle between them determines the intensity value, being maximum when both vectors are parallel.  Top right: Representation of the hierarchical organization of memory types. From the reservoir features of the input pattern $\bm{S}^*$, the \textit{raw memories} $\overline{\mathbf{d}^o}$ (first category) associated to the maximum intensity at their modes $\{I^o\}^M_{max}$ are selected to generate a \textit{tailored memory} $\overline{\mathbf{d}}^\Sigma$ (second category). At the same time, the \textit{tailored memories} are gathered into \textit{memory classes} $c$ depending on the class they belong to (third category).}
\label{fig: Setup}
\end{figure}

In our optical version, illustrated in Fig.\ref{fig: Setup}, the generation of noisy replicas is replaced by a similarity selection: essentially selecting the $M$ modes among $O$ that contain the most similar \textit{raw memories} to the target pattern to be written $\textit{S}^*$. One possibility would be to measure all the $\bm{d}^o$ individually, as previously proposed by the authors in \cite{Leonetti2024}. Nevertheless, this operation is time-, hardware- and computation power- consuming because it requires interferometric measurement and digital matrix inversion to retrieve the the dominant eigenvectors values. Here, we propose an alternative, and energy-efficient approach that resorts to fully analog intensity measurements to measure the degree of similarity between a target pattern we aim to \textbf{write}, the pattern $\textbf{S*}$, and the \textit{raw memories} $\bm{d}^o$. Indeed, we will resort to the fact that intensity is also a proxy of the degree of similitude between the query pattern (here we will present experimentally the pattern to be written $\bm{S}^*$) and the stored \textit{raw memory} $\bm{d}^o$ as $I^o(\bm{S})=|\bm{f}^o \cdot  \bm{S}^\dagger |^2$ (see insight \textit{(iii)} before). 

The ratio between the size of the dictionary ($O$, light modes monitored) and the number of features in the input ($N$ i.e. size of the image to be stored), is driving $PhEL$ effectiveness: intuitively the larger the dictionary, the easier to find, between many random patterns, one similar to our target. The power of the $PhEL$ approach is fully shown in Fig.\ref{fig: Emergent Hamming}a.

Open (Full) dots represent the distance between the selected \text{raw memories} (effectively written \textit{tailored memories}) from the target pattern $\bm{S}^*$. The result is remarkable since, starting from almost chaotic memories, just barely aligned to the target $\left(\left\langle H(\bm{S}^*,\bm{\overline{d}}^m)\right\rangle\sim 0.45\right)$, we obtain 
a \textit{tailored memory} which is essentially aligned to $\bm{S}^*$ $\left(\left\langle H(\bm{S}^*,\bm{\overline{d}}^m)\right\rangle \ < \ 0.02\right)$. This is a clear evidence of the effectiveness of the emergent learning, capable with his ``noise canceling'' averaging effect to transform a set of almost disorganized patterns into an user designed structure, enabling the possibility to extract ``meaning'' from disorder. 

The effectiveness of the \textit{writing} process depends also the number of modes $M$ which have been merged, this dependence is (reported in Figure \ref{fig: Emergent Hamming}d). 

$PhEL$ allows not only to \textit{read} and \textit{store} memories but also enables our SPRC to work as a classifier simply employing \textit{tailored memories} as examples and exploiting insight \textit{(iii)} by demonstrating that aggregated intensity is an analog measurement of the degree of similarity of the stored modes with the query pattern:
\begin{equation}
    I^\Sigma(\bm{S})=\sum_m^{1..M}I^{m}(\bm{S}) =\bm{S} \cdot \mathbb{R}^\Sigma  \cdot \bm{S}^\dagger\sim |\bm{d}^\Sigma \cdot \bm{S}^\dagger|^2
    \label{eq:SPRC Intensity Sigma 2}
\end{equation}

To realize the classification we have first to store examples in our SPRC, in the form of \textit{tailored memories}. For each class $c$ we store $P$ examples (examples index is $p$). For each class then we retrieve the maximal intensity between the examples. This \textit{class aggregated intensity} is thus 

\begin{equation}
  I_{c}(S) = MAX(\{ \bm{I}^{\Sigma,p}(\bm{S})\}_c)
    \label{eq: class aggregated-intensity}
\end{equation}  



Then, similarly to the winner-takes-all rule used in neural networks, the predicted class $c$ corresponds to the class with the highest intensity response $I^{c}(S)$. This process, illustrated in Fig.\ref{fig:classification}a, is equivalent to compare the query with $P\times C$ examples. This aspect will be thoroughly described in the results.

\section*{Results}
This section presents experimental highlights demonstrating how a standard SPRC implements the $PhEL$ paradigm for reading-writing memories and recognizing-classifying queries. The system experimentally validates claim \textit{(iii)}, with output intensity serving as a proxy for similarity between input patterns and stored memories, achieving memory storage and retrieval quality assessment up to 99.9$\%$. Under specific experimental conditions, the system's storage capacity reaches almost infinite capacity (approximately a number of patterns $\mathcal{M} \sim 10^{60557}$), while enabling recognition with an error of around $10^{-4}$ and classification efficiency of 93.5$\%$.

\subsection*{Writing-Reading}
As described above and represented in Fig.\ref{fig: Setup}, the first step to write a memory is to select a combination of modes $o$ which successfully represent the target pattern $\bm{S}^*$. We first present $\bm{S}^*$ on the query layer (the DMD), with phase encoding (using superpixel method \cite{Goorden2014}, see \hyperref[sec:Methods]{Methods}) so that $S_n\in[\pm 1]$. Then we sort the output intensities from the physical repository and finally select the "best" pixels according to intensity. This repository is typically large: employing a camera detector one can associate to each camera pixel one light mode (refer to description of the experimental setup in \hyperref[sec:Methods]{Methods}), thus monitoring tens or even hundreds of millions ($O \sim 10^6$) of light modes (i.e. \textit{raw memories}).

\begin{figure}[ht]
\centering
\includegraphics[width=0.9\linewidth]{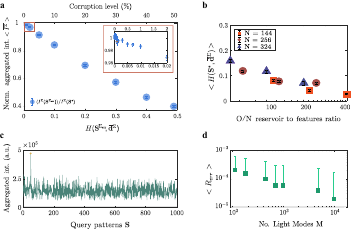}
\caption{\textit{Characterization of the SPRC system.} \textbf{(a)} Memory capacity assessment based on the \textit{normalized aggregated intensity} $\overline{I^{\Sigma}}$ of the query pattern $\bm{S}^{\Sigma_{err}}$ using its \textit{tailored memory} $\overline{\bm{d}}^\Sigma$ as a function of the level of corruption present in the query. Inset zooms the region at very low corruption levels. Orange point sets the level at which the corrupted patterns $\bm{S}^{\Sigma_{err}}$ and the original $\bm{S}^*$ start to be distinguishable. Here we employ bigger input patterns $N$ = 112896 pixels to demonstrate the feasibility of $PhEL$ in the very large $N$ region. \textbf{(b)} Average Hamming distance as a function of the reservoir to features ratio $O/N$ when recovering the \textit{tailored memory} of input pattern $\bm{S}^*$ using $M$ = 250 modes in our system for different $O$. \textbf{(c)} Recognition assessment of specific pattern $\bm{S}^*$ through the \textit{aggregated intensity} of the query patterns $I^{MNI}(\bm{S}^{MNI})$ with respect to the \textit{tailored memory} $\overline{\bm{d}}^{MNI}$ to be recognized. The number of modes used to design the memory is $M$ = 1000. \textbf{(d)} Mean recognition error and standard deviation are calculated respectively by averaging and computing standard deviation over 100 patterns. We report data for the first $10^3$ query probe patterns, while writing is performed with $M= 10^3$} 
\label{fig:miscellaneous}
\end{figure}

To quantify the effectiveness of both writing and reading processes we resorted to various experiments reported in Fig.\ref{fig:miscellaneous}.  We use the MNIST dataset \cite{MNIST} (containing handwritten digits) as input patterns. After the pattern $\overline{\bm{d}}^{\Sigma} \sim \bm{S}^*$ is stored in our optical device, we proceed to the characterization of the writing efficiency (Fig.\ref{fig:miscellaneous}a and its inset) by measuring the aggregated intensity $I^{\Sigma} (\bm{S}^{\Sigma_{err}})$, obtained presenting the pattern $\bm{S}^{\Sigma_{err}}$ containing a corrupted version of the \textit{tailored-memory}, (we flipped randomly a percentage of the $N$ original pattern binary features). Then we monitor quantitatively the average \textit{normalized aggregated intensity}: $\overline{I^{\Sigma}}=<I^{\Sigma} (\bm{S}^{\Sigma_{err}})>/I^{\Sigma} (\bm{S}^{*})$. 

It is possible to see that the $\overline{I^{\Sigma}}$, linearly decreases with the introduced corruption level, further supporting both insight \textit{(iii)} (i.e. that aggregated intensity is a good proxy of the difference between stored and query pattern) and the fidelity of our writing process. 
The orange point in the inset indicates the degree of difference (introduced in the form of pixel flipping corruption) at which the \textit{aggregated intensity} decrement enables to distinguish (in concrete, \textit{aggregated intensity} decreases more than the standard deviation of $\overline{I^{\Sigma}}$) the stored pattern from its corrupted version indicating a writing efficiency of the 0.1$\%$. 

Further evidence of the fidelity is reported in panel Fig.\ref{fig:miscellaneous}b where we compare the average Hamming distance between the stored pattern $\overline{\bm{d}}^{\Sigma}$ (retrieved with the Complete Coupling Method \cite{leonetti2023reference}, more in \hyperref[sec:Methods]{Methods}) and the target pattern $\bm{S}^*$. Here, the parameter driving our $PhEL$-based machine is the ratio between the dictionary size ($O$, the number of camera pixel, i.e. the number of available \textit{raw memories}) and the length of the word to be written ($N$, the number of features in $\bm{S}^*$, i.e. the number of employed DMD pixels). It is possible to note that the average Hamming distance lowers while increasing ${O}/{N}$ corresponding to an increasingly successful writing process (in this case, $M$ = 250 light modes). These results experimentally demonstrate the theoretical simulation in Fig.\ref{fig: Emergent Hamming}a. They quantitatively confirm the ability of the system to generate a fauthful \textit{tailored-memory} from an input pattern. The Hamming distance in experiment is higher (lower similitude between input pattern and stored memory) than in simulations due to unavoidable experimental noise sources.




Using the memory capacity of the system, $PhEL$ can be used to recognize patterns optically relying only on  $I^\Sigma (\bm{S})$ comparison. Fig.\ref{fig:miscellaneous}c accounts for the process of  recognizing a pattern among a large set of samples ($10^4$ patterns, where only the first $10^3$ are shown).  In the measurement, we proceed as follows: we write a single pattern taken randomly from the MNIST repository ($\bm{S}^{MNI*}$) with the process described above, realizing the memory pattern $\overline{\bm{d}}^{MNI}$ with $M$ = 1000. Then we propose a set of $10^4$ patterns randomly extracted from MNIST training dataset $\bm{S}^{MNI}$ and measure the \textit{aggregated intensity} $I^{MNI} (\bm{S}^{MNI})$. It is possible to acknowledge a clear spike with a yellow dot indicating that the presented pattern is recognized as the one originally written in \textit{tailored memory}. 

Fig.\ref{fig:miscellaneous}d accounts for the mean error when obtained in this recognition process ($10^4$ patterns presented). As shown already in simulations, the recognition error is reduced as the \textit{tailored memory} are constructed with a larger number of selected \textit{raw memories}, $M$.



Finally, one can estimate the maximum capacity of optical operators/ \textit{tailored memories} ($\mathcal{M}$) which can be installed in an SPRC in our experimental conditions. This number of possible \textit{tailored memories} $\bm{d}^ {\Sigma}$ stored in the system corresponds to the combinations relative to the number of available output features of the physical reservoir $O$, the number of $M$ \textit{raw memories} used to design the \textit{tailored memory} and, finally, the size of the input pattern $N$. In our specific experimental conditions, ($O$ = $10^6$, $M$ = 250 and $N$ = 324) $\mathcal{M} \sim 10^{60557}$.

\begin{equation}
    {O\choose M} = \frac{O!}{M!(O-M)!} = {10^6 \choose 250} \sim 10^{60557}
\end{equation}

\subsection*{Classification with analog training}
Once we have demonstrated the capability of writing-reading memories, one can perform classification of unknown queries. From the hierarchical organization of the memories, the system is trained by analog generation of \textit{tailored memories}. The decision layer is constructed by grouping the stored \textit{tailored memories} into \textit{memory classes} without additional computational cost (see Fig.\ref{fig: Setup}). 

We tested the system on the well-known machine learning benchmark, the MNIST dataset \cite{MNIST}. In this experiment, we used a total of 5410 randomly selected images which are organized into C = 10 classes (digits from 0 to 9), in a balanced fashion. Consequently, we stored 4410 \textit{tailored memories} from the training set ($P = 441$ per each class)  and the rest of 1000 patterns for validation (100 patterns per category).

\begin{figure}[ht]
\centering
\includegraphics[width=\linewidth]{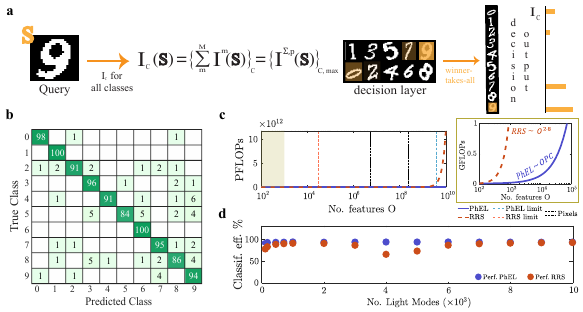}
\caption{\textit{Classification technique and results.} \textbf{(a)} Scheme of the classification flow. The \textit{aggregated intensity} $I^\Sigma$ of each $p$ \textit{tailored memory} is calculated and the maximum value is considered the output of the class $c$. The decision layer is composed of 0 to 9 classes. The highest value in the decision is the classification output. \textbf{(b)} Confusion matrix of the classification of 1000 patterns of the MNIST dataset using $M$ = 1000 light modes. \textbf{(c)} Simulation of the computational complexity in FLOPs of classification training, using P = 441 samples per class of size $N$ = 15 x 15 for C = 10 digit categories, applying the $PhEL$ (blue line) and the $RRS$ approach \cite{Saade2016} (red line). Vertical dashed lines set set the limits of computational resources using $RRS$ (light red) at $O$ = $3\cdot 10^4$, $PhEL$ (light blue) at $O$ = $4\cdot10^9$ and the average number of pixels of our array detector $O = 5.6\cdot10^6$ (bold black) and the highest number of pixels of a commercial array detector (BASLER) $O = 1.52\cdot10^8$ (light black). \textbf{(d)} Classification efficiency as a function of the number of light modes selected: to generate the \textit{tailored memory} in the optical system ($PhEL$, blue dots) or as features to be introduced in the digital training ($RRS$, red dots).}
\label{fig:classification}
\end{figure}

The classification output layer for the unknown query pattern $\bm{S}$ is conformed by deriving the \textit{aggregated intensity} of each class $c$, $I_c$ (see Equation \ref{eq: class aggregated-intensity}), by deriving the maximum \textit{aggregated intensity} $I^\Sigma$ calculated for $\bm{S}$ from physical measurements as represented in Fig.\ref{fig:classification}a. The final class is selected by comparing the analog retrieval of $I_c$ among all classes $C$, following the winner-takes-all rule for the decision. For detailed information on the training and validation procedure, refer to \hyperref[sec:Methods]{Methods}. 

Fig.\ref{fig:classification}b presents the confusion matrix that synthesizes the classification results and demonstrates a categorical recognition efficiency of 93.5$\%$ (using $M$ = 1000 modes). The dependence of classification performance with different numbers of light modes is shown in Fig.\ref{fig:classification}d. In the low $M$ region ($M$<120), the SPRC based on $PhEL$ shows a classification performance higher than 91 $\%$. 

The potential of $PhEL$ is supported by Fig.\ref{fig:classification}c and Fig.\ref{fig:classification}d, where we compare it with the (simulated) $RRS$ approach \cite{Saade2016}, a very popular photonic computation approach for SPRC, in the levels of computational performance and inference capacity. Both approaches have in common the use of speckle patterns applied to the classification of images. At difference from $PhEL$, the $RRS$ technology 
requires to build a digital layer of parameters, applied to the measured intensity of the camera pixels, which transforms random projectors into a classifier device but requiring a big computational effort \cite{Saade2016}. 

Translating this into computational complexity, see \hyperref[sec:Methods]{Methods} for deeper details, we estimate the number of floating point operations, for both approaches depending on the number of employed light modes. In specific, Fig.\ref{fig:classification}c demonstrates the potential of analog training versus $RRS$ counterpart. Computational cost for $RRS$ is strongly affected by digital inversion of huge matrix ($O\times O$), meanwhile $PhEL$ only requires a computing cost  linearly depending on $O$. The shadowed region is zoomed in the panel on the right and shows the skyrocketing cost of $RRS$ (red dashed curve) with respect to $PhEL$ (blue line). Vertical bars, in the main panel, denote the limits for both approaches ($RRS$ and $PhEL$ in light red and blue, respectively) and black lines indicate the available light modes $O$ (camera pixels) for current camera in our experiments (bold line) and the camera with the highest resolution of the same series (light line). In addition, Fig.\ref{fig:classification}d presents a better classification efficiency ($PhEL$ = 93.5$\%$ vs 91.7$\%$ for $RRS$).
 
\section*{Discussion}
We demonstrate $PhEL$, a method that transforms the triad of a DMD, a disordered medium, and a camera sensor into a versatile optical computing platform. $PhEL$ can store patterns, use \textit{tailored memories} as optical comparators, recognize queries, and classify unknown patterns using stored examples.

Our system exhibits several unique characteristics. \textit{First}, it can load an effectively unlimited number of optical operators that function as analog comparators. 

\textit{Secondly}, installing a \textit{tailored memory} is extremely simple: one only needs to present the pattern to be stored and record the indices of the camera pixels (or light modes) that produce maximal intensities. This writing process leverages \textit{raw memories}  --- random information embedded in the field-transmitted vectors of each light mode---which the system exploits in an emergent manner.

This mechanism forms the core --- and the counterintuitive insight --- of Photonic Emergent Learning. We show that the system can exploit disorganized patterns, only loosely resembling the target pattern, to produce an almost error-free memory. This error reduction is a non-trivial effect, mediated by the matrix eigenvalue operator and related to how memories and attractors emerge in shallow Hopfield networks from corrupted replicas of an archetypal pattern.

As a \textit{third} point of uniqueness, tailored memories also function as optical comparators, achieving a co-localization of memory (storage) and operator (optical comparator).

This co-localization enables the \textit{fourth} point: because pattern comparison and recognition are performed primarily in the analog domain, a significant portion of the computational burden is shifted to the optical layer. Consequently, when patterns are used as examples to train a classification device, the computational cost is dramatically reduced --- from scaling as \( O^{2.8} \) to scaling linearly (\( O \)) with the number of light modes or camera pixels --- while simultaneously offering higher efficiency in the \textit{low-number-of-light-modes} regime.

The \textit{fifth} unique feature is the intrinsic simplicity of the approach: we can deliver fully functional and precisely designed tasks without requiring optical simulations, hardware fabrication, or the addition of a digital processing layer.

We are not proposing a simple platform for computation, but a new paradigm that transfers the computational burden onto the optical layer. Unlike other approaches, we strategically exploit the uniqueness of light--matter interactions and the abundance of mesoscopic disorder in an innovative manner, opening a new avenue for future optical deep computation.

\section*{Materials and Methods}
\label{sec:Methods}

\subsection*{Experimental set-up}
The system is implemented by using a continuous monochromatic laser at 532 nm (Azure Light) that uniformly illuminates a DMD, able to spatially encode digital information on the light beam by modulation of amplitude composed of 1024 x 768 micromirrors and a micromirror pitch of 13.7 $\mu m$ (ViALUX, V-7000), see Fig.\ref{fig: Setup}. The light beam carrying the signal is then focused on a random medium by a 4-f system (LA–1708-A and AC254-060-A-ML). Here, the medium is a spray-painted layer on a standard glass microscope cover slip. The white paint (Citadel colour) was applied in two layers by deposing paint during 3 seconds of exposition at a 30-centimetre distance. The transmitted light is then collected on the far side by a second lens (THORLABS, TL2X-SAP) and a tube lens (THORLABS, TTL100-A) followed by a linear polarizer to measure the speckle pattern by a standard monochrome CCD camera (BASLER a2A2600-64umBAS). 

The writing process of a \textit{tailored memory} $\mathbf{d}^{\Sigma}$ consists of coding the desired pattern $\bm{S}$ in the input coherent light using a DMD. The DMD is an amplitude modulation device whose micromirrors can be set at two positions 0/1. One state sends the light into the optical path of the disordered medium, and the other state reflects light towards a beam blocker. Despite this, we applied a superpixel method \cite{Goorden2014} that allows the DMD to introduce phase information in the wavefront of coherent light. This method further encodes the input image by transforming each pixel into a Hadamard pattern (+1 or -1, see Fig.\ref{fig: Setup}) depending on its value over a set threshold. In all the experiments, a superpixel method of 2 x 2 is employed.  

After passing through the diffusive medium, the related speckle pattern is recorded using a camera. Note that to simplify the realization of the experiment, we operate in the configuration in which each mode $o$ corresponds to a single sensor. As we employ a camera to measure $I^o$, the configuration of one mode per pixel is obtained by properly tuning the optical magnification (in our case x1), that is, matching the speckle mode size and the pixel size. The ROI set in the detection is $O$ = $1000^2$, being the total number of the available modes.

As previously defined, the \textit{tailored memory} $\mathbf{d}^{\Sigma}$ corresponding to the input pattern $\bm{S}^*$ is extracted by taking the \textit{raw memories} at the index position $o$ of the most $M$ intense pixels $I^o$ among the total $O$ modes. This pixel selection is performed by applying the Quickselect algorithm \cite{Quickselect} to the output image of the reservoir, which finds the $M$ largest elements in an unordered list.

\subsection*{Emergent Learning with binary phase patterns}
The parallelism between the Hopfield memory model and optics has some limitations. Notably, the vector $\bm{f}^{o}$ is composed of complex numbers, whereas in the traditional Hebbian framework, the pattern to be stored is a boolean (real) vector. Indeed, the mode $o$ possesses a complex-valued (non-Boolean) memory $\bm{f}^{o}$, which by construction is the unique eigenvector of the \textit{optical-synaptic matrix} $\mathbb{F}^o$.

As one is generally interested in storing information in the form of binary bits, we decided to employ a  digital micromirror device (DMD) in the phase-only mode {($\phi_n\in [0 - \pi]$)},  which is equivalent to a real modulation of the input amplitude   with binary multiplicative  coefficients $S_n =e^{j\phi_n} \Rightarrow S_n \in [-1,+1 ]$. This approach introduces a rank-2 degree of nonlinearity between input and detection (phase and intensity), at the same time increasing the resilience of the output with respect to small phase fluctuations.

In the case the input modulation is real, we note that the imaginary part of the \textit{optical-synaptic matrix} $\mathbb{F}^o$ does not play any role in determining the intensity (which is also real), thus has to somehow vanish away. Indeed, the shape of a single element of the \textit{optical-synaptic matrix} is such that the transmitted fields are complex fields:  $f^{o}_n=a^o_n+jb^o_n$ or in vectorial form $\bm{f}^{o}=\bm{a}^o+j\bm{b}^o$ then 

\begin{eqnarray*}
    \mathbb{F}^o_{n,l} =f^{o}_n\otimes f^{o \dagger}_l=a^o_na^o_l+ b^o_nb^o_l +i(a^o_nb^o_l- b^o_na^o_l)   \\
    \mathbb{F}^o_{l,n} =\mathbb{F}^{o\dagger}_{n,l} = f^{o}_l\otimes f^{o ^\dagger}_n=a^o_na^o_l+ b^o_nb^o_l -i(a^o_nb^o_l- b^o_na^o_l)\\
    \label{T_Single_Element}
\end{eqnarray*}

being $\mathbb{F}^o$ an hermitian matrix. In the computation of the intensity, the imaginary part of the $\mathbb{F}^o$ matrix sums to zero, as expected for a real and positive intensity (which is also real, see Equation \ref{eq:SPRC Intensity nu}). Indeed, the matrix $\mathbb{R}^o$, the  real part of $\ {\mathbb{F}^o}$ has the property 

\begin{equation}
I^o(\bm{S}) =\bm{S} \cdot \mathbb{F}^o  \cdot \bm{S}^\dagger = \bm{S} \cdot \mathbb{R}^o  \cdot \bm{S}^\dagger.
\label{Real_Int}
\end{equation}

We note that $\mathbb{R}^o$ is a bi-dyadic matrix

\begin{equation}
\mathbb{R}^o=\bm{a}^{o}\otimes\bm{a}^{o\dagger} +\bm{b}^{o}\otimes\bm{b}^{o\dagger}  
\label{eq:RRRshape}
\end{equation}

which, by construction, is characterized by two eigenvectors $\bm{a}^o$ and $\bm{b}^o$. Among these, we define ${\bm{d}}^o$ as the dominant eigenvector, i.e., the one with the highest eigenvalue $\lambda_d^o$.

By applying the input $\bm{S}={\bm{d}}^o$ at our input layer, we are maximizing the intensity at the mode $o$: $I^o({\bm{d}^o})$  maximize it with respect to any possible real-only modulation. Thus, also in the case of real modulation, mode $o$ contains a random and uncontrolled memory (\textit{raw memory} embodied by the vector ${\bm{d}}^o$). If binary phase (amplitude) modulation is employed, then the input maximizing the intensity is instead $\overline{\bm{d}}^o=sign( {\bm{d}^o})$. 

\subsection*{Parallelism with the Hebbian Learning}

Interestingly, by considering that in the first approximation $\mathbb{R}^m\sim \bm{d}^m \otimes\bm{d}^{m\dagger}$ , the  \textit{tailored optical-synaptic matrix} then has the shape

\begin{equation}
\mathbb{R}^\Sigma\sim \bm{d}^1 \otimes\bm{d}^{1\dagger}+\bm{d}^2 \otimes\bm{d}^{2\dagger}+...+ \bm{d}^M \otimes\bm{d}^{M\dagger} 
    \label{eq:RSigmaHopfield}
\end{equation}

which is the same shape for a synaptic matrix containing $M$ \textit{raw memories} in the Hopfield model \cite{hopfield1982neural, amit1985storing}.

\subsection*{Characterization of $PhEL$: simulations and experiments}

The Hamming distance $H(\bm{A},\bm{B})$ refers to the number of positions at which two vectors $\bm{A}$ and $\bm{B}$ of the same length ($N$) differ \cite{Hamming2016}. It is a metric used in computer science to measure dissimilarity between strings.

\begin{equation}
    H(\bm{A},\bm{B}) = \frac{\bm{A}\cdot\bm{B}}{N}
\end{equation}

\begin{enumerate}[label=(\roman*)]
\item $H(\bm{A},\bm{B}) = 0.5$ shows that $\bm{A}$ and $\bm{B}$ are uncorrelated. 

\item $H(\bm{A},\bm{B}) = 0$, when all vector elements are equal. 

\item $H(\bm{A},\bm{B}) = 1$, when all vector elements are distinct. 
\end{enumerate}

In the simulations of Fig.\ref{fig: Emergent Hamming}, 
we report the results of numerical simulations to provide insight into the photonic emergent learning process. To realize the simulations, following the analysis also presented in [20,27], 
we initialized the transmitted field vector (representing the fields after the disordered medium, in absence of modulation, i.e. $\bm{S}=\mathit{\bm{I}}$, with $\mathit{\bm{I}}$ an identity vector), as a set of randomly extracted real and complex coefficients ($a_n,b_n$) with mean 0 and variance 1. The number of modes (camera pixels)  has been set to  $O=5 \cdot10^5$. The intensity in each mode for any probe pattern $\bm{S}$ can be retrieved by exploiting Equation \ref{Real_Int}. With this simple approach, we can perform all the tasks required in Fig.\ref{fig: Emergent Hamming}. To retrieve the stored \textit{tailored memory} ($\overline{\bm{d}}^\Sigma$), we can both retrieve the eigenvector of the \textit{tailored optical-synaptic matrix} ($\mathbb{R}^{\Sigma}=\sum^M\mathbb{R}^{m}$) or perform a numerical iterative wavefront shaping in which \textit{aggregated intensity} (to be optimized) is computed by summing the individual intensity of each mode ($I^\Sigma(\bm{S})=\sum^MI^m(\bm{S})$). At each iteration step, we flip one random input $S_n$ and accept the change only if $I^\Sigma(\bm{S})$ increases. Four consecutive iterations over all pixels ($4N$ steps) ensure full optimization.

In the memory capacity experiment, we study the precision of the system in distinguishing between two input patterns that differ by an increase in the percentage of corrupted pixels. In this case, we used an input pattern size of $N$ = 112896 to assess differences at low levels of corruption. For that, a preprocessing stage is performed that rescales the cropped 24 x 24 MNIST images followed by binarization, thresholding, and the application of the superpixel method. As the active are of the DMD has been changed up to 672 x 672 micromirrors, after superpixel method, a telescope is added before the DMD to expand the beam area fully covering the DMD. A total of 10 random digits from MNIST are tested. For each digit, its \textit{tailored memory} $\mathbf{d}^{\Sigma}$ is generated and the query patterns $\bm{S^{\Sigma_{err}}}$, which are corrupted at different levels ranging from 0.01$\%$ up to $50\%$, are projected into the DMD. Each bunch of measurements comprises a set of 101 projections, the original pattern $\bm{S}^{*}$ followed by 100 repetitions affected by a specific percentage $\bm{S}^{\Sigma_{err}}$. Corrupted versions of the input images are created after binarization in preprocessing by flipping the values 0/1 to the corresponding number of pixels. For each bunch of measurements, the average value of the \textit{normalized aggregated intensity} $\overline{I^{\Sigma}}$ is calculated averaging the \textit{aggregated intensity} of the replicas $I^{\Sigma} (\bm{S}^{\Sigma_{err}})$ and normalizing by the $I^{\Sigma} (\bm{S}^{*})$. 

Recognition is performed using MNIST preprocessed pattern, as previously explained, of size $N$ = 225. Firstly, $P = 10^4$ probe patterns are shown to the reservoir to retrieve their associated \textit{tailored memories} $\mathbf{d}^{\Sigma}$ ($M$ = 1000). Afterwards, 100 query patterns, belonging to these probe patterns, are projected one by one, and the final \textit{aggregated intensity} curve $I^{\Sigma}$ is calculated using the stored \textit{tailored memories} of the probe patterns in each projection. To assess the recognition capacity of the system to correctly detect the input pattern among the probe patterns, the figure of merit shows the number of detections ($\#$ detection) that surpass a threshold of $18\sigma_{I}$ normalized by $P$. Then, the mean recognition error averaged over the 100 tests and its standard deviation are calculated. 

\begin{equation}
    <R_{err}> = \frac{\# detection}{P}
\end{equation}

To experimentally study the influence of the size pattern and the available light modes dependence to faithfully generate a \textit{tailored memory}, we generate \textit{tailored memories} and afterwards we try to read them from the system. As measurement conditions, all possible combinations of three different input pattern sizes $N$ = $144$, $256$, $324$, and three different ROI sizes $O$. Recovery of the \textit{tailored memory} from the system, called the reading process, is experimentally performed using the Complete Coupling Method \cite{Leonetti2024, leonetti2023reference}. Using the same procedure, we retrieve the saved \textit{tailored memory} from the system for an input pattern. To quantify the fidelity of the input pattern and its stored \textit{tailored memory}, we use the Hamming distance between the recovered \textit{tailored memory} and the input pattern $\bm{S}^*$. 

\subsection*{Classification procedure}
Classification performance is evaluated using the MNIST handwritten digit dataset \cite{MNIST}. Pre-processing should be addressed before projecting the pattern with the DMD. Each digit in the MNIST dataset can be seen as a 28 x 28 array of integers between 0 and 255. We first adjust the size of the images by cutting the dark edges to ensure a larger area of interest when displaying the pattern and resizing the image to $N$ = 225. Then, we quantize the grey levels between 0 and 1 to further apply the superpixel method. The final correspondence between the input image and the projected pattern is 30 x 30 DMD micromirrors. 

Training consists of writing the associated \textit{tailored memories} from 4410  digits of the MNIST repository using $M$ = 1000 modes. Then these \textit{tailored memories} are grouped into 10 \textit{memory classes} $c$, with no additional computational overhead, whose categories represent digits 0 to 9 and constitute the output layer. The DMD is set to work triggered to the camera frame rate. The system essentially performed optical training in just 33 minutes, comprising speckle detection, \textit{tailored memory} generation and categorization.

To validate the $PhEL$-based classification performance, 1000 patterns are measured per category. Recalling that training \textit{tailored memories} are grouped into their categories, for each input query pattern $\bm{S}$, the speckle pattern that emerges from the reservoir is used to evaluate the \textit{aggregated intensity} associated for each class. Hence, 4410 simultaneously \textit{aggregated intensities} are calculated. The highest \textit{aggregated intensities} at each class gives the output vector of this decision layer. The classification result comes applying the winner-takes-all rule, this is from the maximum of the output. Fig.\ref{fig:classification}(a) exemplifies the classification process. The total time to measure the total number of query patterns and perform their classification was 3 minutes.

To properly compare the results of $RRS$ and $PhEL$ approaches  in classification efficiency, the $RRS$ process must be simulated. For that, we trained a final linear ridge regression layer using simulated speckle patterns after passing from a diffusive medium as described in \cite{Saade2016}, where the input pattern $\mathbf{U_{t}}$ is non-linearly projected in a higher dimensional space:

\begin{equation}
    \mathbf{X_{t/v}} = f(\mathbf{W}\mathbf{U_{t/v}})
\end{equation}
where \textbf{t} stands for training and \textbf{v} for validation. 

We used the same conditions of the $PhEL$ experiment, as well as, the same input patterns $\mathbf{U_{t}}$ and labels $\mathbf{Y_{t}}$ of MNIST \cite{MNIST} as in the $PhEL$ experiment (P = 441 per class of size $N$ = 15 x 15 for $C$ = 10 digit categories). The "measured" speckle patterns ($\mathbf{X_{t}}$) are retrieved by multiplying the input patterns by the simulated transmission matrix $\mathbf{W}$ relative to a diffusive medium (an i.d.d. random complex matrix) \cite{Popoff2010,Saade2016} and after applying the modulus function, corresponding to the intensity detection of a camera. The number of features of the speckles $O$ used for training ranges from $10^2$ to $10^{10}$. The classification efficiency is evaluated using the same test patterns as in the $PhEL$ experiment ($\tilde{P} = 100)$ for coherence. The inference output $\mathbf{Y_{out}} \in \mathbb{R}^{(\tilde{P}C)xC}$ is calculated on the simulated speckles ($\mathbf{X_v}$) and the trained weights $\mathbf{Q}$. 

\begin{equation}
    \mathbf{Y_{out}} = \mathbf{X_v}(\mathbf{X}
    _{t}^\dagger \mathbf{X_{t}} + \alpha \mathbf{I_O})^{-1} \mathbf{X^\dagger_{t}} \mathbf{Y_t} = \mathbf{X_{v}}\mathbf{Q} 
    \label{eq:RR}
\end{equation}

where $\mathbf{X_{v}} \in \mathbb{R}^{(\tilde{P}C)xO}$, $\mathbf{X_{t}^\dagger} \in \mathbb{R}^{Ox(PC)}$, $\mathbf{X_{t}} \in \mathbb{R}^{(PC)xO}$, $\mathbf{I_{O}} \in \mathbb{R}^{OxO}$, $\mathbf{Y_{t}} \in \mathbb{R}^{(PC) x C}$ and $\alpha \in \mathbb{R}$. 

The computational complexity is also theoretically estimated both for $PhEL$ and $RRS$ methods. We derived the number of computational calculations that each approach needs focusing on training, as it is the most expensive task. Calculations use the selected features of the speckle pattern for all training patterns $\mathbf{X_{t}}$.

On the one hand, the SPRC based on $PhEL$ uses the measured features of the physical reservoir output of $P \cdot C$ total \textit{tailored memories} for training. These $M$ features are selected by applying the maximum intensity mode criterion, see Equation \ref{eq:SPRC Intensity Sigma 2}. Quickselect algorithm \cite{Quickselect} finds these largest $M$ values for all total\textit{ tailored memories} with a cost of

\begin{equation}
    (O+Mlog_2M)PC
\end{equation}

being $O$ $(>> M)$ the total ROI of the camera. Then, these \textit{tailored memories} are gathered into \textit{memory classes} at no additional computational expense. 

On the other hand, the $RRS$ method relies on solving the equation of the ridge regression problem, previously shown in Equation \ref{eq:RR}, using all the available modes. There, three matrix multiplications and a matrix inversion occur needing approximately the following calculations in FLOPs:  

\begin{align}
        Multiplications: O^{2.8} + O^2C(P+1) + OPC^2 \\
        Additions: O^2(PC+1)- O^2 + OC(PC-2) + O \\
        Total: O^{2.8} + 2O^2C(P+1) - O^2 + 2OC(PC-1) + O
\end{align}

\section*{Author contributions statement}

S.P.G and M.L. conceived the experiments, S.P.G. conducted the experiments, S.P.G. analyzed data, G.G. and S.P.G. performed simulation for $RRS$ and M.L. and S.P.G. for $PhEL$. S.P.G. and M.L. wrote the manuscript. All authors discussed the results and commented on the manuscript. 

\section*{Conflict of Interest}

Authors declare no conflicts of interest.

\bibliography{Paper_overleaf.bib}

\end{document}